# Inferring Economic Condition Uncertainty from Electricity Smart Meter Data

Haoqi Qian[a,b], Zhengyu Shi[c], Libo Wu[d,e,1]

[a]Institute for Global Public Policy, Fudan University, Shanghai 200433, China
[b]LSE-Fudan Research Centre for Global Public Policy, Fudan University, Shanghai 200433, China
[c]School of Data Science, Fudan University, Shanghai 200433, China
[d]School of Economics, Fudan University, Shanghai 200433, China
[e]Institute for Big Data, Fudan University, Shanghai 200433, China

**Abstract:** Inferring the uncertainty in economic conditions is significant for both decision makers as well as market players. In this paper, we propose a novel approach to measure the economic uncertainties by using the Hidden Markov Model (HMM). We construct a dimensionless index, Economic Condition Uncertainty (ECU) index, which ranges from zero to one and is comparable among sectors, regions and periods. We used the daily electricity consumption data of more than 18,000 firms in Shanghai from 2018 to 2020 to construct the ECU indexes. Results show that all ECU indexes, whether at sectoral or regional level, successfully captured the negative impacts of COVID-19 on Shanghai's economic conditions. Besides, the ECU indexes also presented the heterogeneities in different districts as well as in different sectors. This reflects the facts that changes in the uncertainty of economic conditions are mainly related to regional economic structures and targeted regulatory policies faced by sectors. The ECU index can also be readily extended to measure the uncertainty of economic conditions in various realms, which has great potentials in the future.

**Keywords:** Economic condition uncertainty, Electricity smart meter data, Hidden Markov Model, COVID-19, Firm-level analysis

---

# 1 Introduction

Inferring economic conditions timely is crucial for both market players and decision makers to understand the current economic status and make their plans. Typically, decision makers and researchers use specific variables to construct quantitative indicators to reveal and analyze the temporal and spatial economic conditions (J. Song, 2018; Yu & Du, 2019). For example, statistical indicators such as gross domestic product (GDP), total export-import volume, and gross investment that are released by national statistical agency are often used to indicate economic situations from different aspects (de Lange et al., 2014; Shin et al., 2018). Appropriately designed economic condition indexes can provide solid evidence for decision makers to make proper policy adjustments under the economic fluctuations (Holm & Østergaard, 2015; Obschonka et al., 2016; Sensier et al., 2016). However, the changes of international situations, policies, market sentiment and other unobservable situations may lash the economic conditions so that there are still difficulties to measure uncertainties accurately (Bloom, 2014).

In the face of economic uncertainty, traditional indicators have two main disadvantages in practice. Firstly, most of the existing economic condition indexes are constructed with absolute values of macroeconomic variables. Although economic activities and real business cycles can be forecasted by using complex modelling techniques based on traditional indicators (Ng & Wright, 2013), these indexes may often lag thus can hardly identify prompt responses in economic conditions when facing unexpected exogenous shocks such as earthquake, flood, epidemic, war, etc. As a result, impacts of these unexpected events on societies and economies are usually underestimated (Markhvida et al., 2020; L. Zhou & Chen, 2021). Policy makers need new indexes that can provide effective information on economic uncertainties to facilitate the economic recovery contradictorily (Peters, 2021; Tierney, 2012). Secondly, the existing indexes are typically released with limited frequency, such as monthly, quarterly, or even annually. This drawback reduces their application of assisting policy makers to make prompt decisions in policy adjustments. Therefore, it is necessary to find a way to construct new indices with both high-resolution and high-frequency in order to make macro-control more effectively.

To avoid the aforementioned drawbacks, this paper attempts to use smart meter data to construct a novel quantitative index through a bottom-up approach that can reveal the economic conditions under uncertainty. We call it Economic Condition Uncertainty (ECU) index. Although many scholars have already used various data sources such as night light (Gibson et al., 2020), newspaper (Baker et al., 2016), and mobile phone data (Dong et al., 2017) to estimate the national or regional economic conditions, the uniqueness of electricity smart meter data makes it more accurate and prompt in capturing the dynamics of economic activities. In addition to the fact that electricity is one of the essential input factors used in modern productions and operations, smart meter data can provide timely and accurate record of the electricity consumption of various units. Therefore, if the long-term stable correlation between electricity demand and macroeconomic prosperity could persist in the short-term and be applied to micro market entities, more accurate economic forecasting and nowcasting could become feasible. However, existing studies using electricity smart meter data in forecasting economic conditions have not fully identified the heterogeneities of micro-level mechanisms that drives the market entities' electricity use. To avoid getting mudded by the inability to fully capture various factors that affect the electricity demand of micro market entities, this paper attempts to model the individual firm's dynamic behavior by using the Hidden Markov Model (HMM) with two unobserved regimes, one representing prosperous regime, and another representing recessionary regime. After the estimation procedure, the unobserved regimes of every firm will be assigned a probability that determines whether the firm suffers from the vicious lash and will be used to construct the ECU index later. We used daily electricity consumption data from more than 18,000 firms in Shanghai from November 2018 to May 2020 to conduct the empirical analyses. Results showed that the ECU indexes constructed in this paper can successfully capture economic fluctuations at both regional and sectoral level.

The rest of this paper is organized as follows. Section 2 reviews the existing relevant literature. Section 3 is the research design that introduces the HMM and methods for constructing the ECU index. Section 4 describes the data used in this study. Section 5 shows the results of different ECU indexes. Section 6 concludes the paper.

## 2 Literature Review

### 2.1 Using proxy variables to measure economic condition uncertainties

In order to characterize the economy conditions, governments and organizations have published various quantitative indexes. For a long time, it is very popular to use composite indexes derived from existing statistical indicators to infer economic conditions worldwide. For instance, the World Bank publishes a series of indexes called the World Development Indicators (WDI) annually, which covers more than 200 countries and regions (WB, n.d.). WDI has an economy column that combines GDP, investment, international trade, central government budgets, prices, etc. to measure the macroeconomic performance specifically; this column has been used to describe the national economic uncertainties (Goel & Ram, 2013). Another typical advanced statistical index is the Economic Conditions Index (ECI) released by the Federal Reserve Bank of the United States (FED, n.d.). This index attempts to measure economic activities in metropolitan areas and is constructed by using a dynamic factor model with 12 traditional indicators including unemployment rate, per capita income, gross metropolis product, etc. (Arias et al., 2016). ECI is a representative index among other similar indexes that are used to measure economic prosperities (Koop et al., 2020; Lasarte-López et al., 2020). Similarly, China National Bureau of Statistics and the People's Bank of China jointly publish the China Macro-economic Prosperity Index (NBSPRC, n.d.) and Enterprises Prosperity Index (PBC, n.d.) respectively to track and forecast the economic conditions.

The aforementioned statistical composite indexes have high dependence on the quality of statistical indicators and surveys, which makes them less satisfactory in terms of timeliness, flexibility and cost performance. Besides, they focus more on the economic reality, but ignore the uncertainties embodied in the economic conditions. The rise of big data in the past decade provides researchers with opportunities to measure the economic conditions as well as the associated uncertainties through different perspectives (Bok et al., 2018). For example, an ongoing project, namely Economic Tracker, uses high frequency economic activity data to track U.S. economy conditions including areas including business, employment, education and public health (Chetty et al.). For other data sources, night light data has gained increasing popularity in measuring regional economic activities, especially for those underdeveloped regions (Gibson et al., 2020; Henderson et al., 2012). Although new night lights data sources such as VIIRS data can provide more unbiased estimates for economic activities compared to traditional DMSP data with a monthly basis, these practices still attract many ongoing debates to be resolved in the future (Gibson et al., 2021). Similar satellite data have also been used in economics to infer conditions in specific areas such as agricultural developments, traffic conditions (Donaldson & Storeygard, 2016), predicting poverties (Jean et al., 2016), etc. Some big data applications also include using social media data (Norbutas & Corten, 2018), mobile phone data (Dong et al., 2017), etc.

Furthermore, in order to track policy-related conditions, a news-based index called Economic Policy Uncertainty (EPU) is proposed and has gained popularity among academia and policymaking areas (Baker et al., 2016). The EPU index detects uncertainties of economic policies by extracting information from main stream newspapers. Such method has been greatly extended to larger scale datasets (Huang & Luk, 2020). More advanced machine learning techniques have also been used to improve the EPU index (Azqueta-Gavaldón, 2017). There are lots of derived indexes based on EPU, such as World Uncertainty Index (WUI) (Ahir et al., 2018; IMF, n.d.), World Trade Uncertainty (WTU) (Ahir et al., 2019) and Geopolitical Risk (GPR) index (Caldara & Iacoviello, 2018). However, these indexes only capture uncertainties from specific sides, that is, policy-related economic uncertainty in EPU and WUI, trade uncertainty in WTU and geopolitical uncertainty in GPR.

Compared to the above applications that use large-scale economic proxy data from different sources, electricity consumption data is often considered as the most appropriate indicator of the economic conditions (Liu et al., 2018; Yalta, 2011). Since electricity consumptions are closely related to the production activities (Zhang et al., 2017), the economic condition fluctuates can be directly reflected by the changes of electricity consumption (Carvalho et al., 2021; Lee et al., 2021). The existing literature has demonstrated the statistical relationships between electricity consumption and economic activities (Amisano & Geweke, 2017; Bah & Azam, 2017; Beyer et al., 2021; Shahbaz et al., 2014, 2017; Sun & Anwar, 2015). With the rapid development and deployment of smart meter technologies, large volume of electricity consumption data can be readily collected in real-time along with high data quality. These features enhance the value of electricity smart meter data in providing more accurate and timely information of economic conditions (Zhou et al., 2019). During the outbreak of COVID-19, high-frequency electricity consumption data has made great contributions towards understanding the impacts and recovery procedures of different countries' economies (Carvalho et al., 2021; Krarti & Aldubyan, 2021). For China, the Resumption Power Index (XinhuaNet, 2020) released by the State Grid Zhejiang Electric Power Co. Ltd. considers the resumption of regional enterprises on the basis of their previous and current electricity consumption. It once became an important indicator to measure the trends of the economic resumption after the COVID-19. However, the Resumption Power Index may not be the most suitable indicator of economic recovery. The increase of electricity consumption does not always mean that the economic condition has become more stable; on the contrary, the resumption of various industries is not synchronized, and some industries are even still restricted, which makes the supply of raw materials in the upstream and sales channels in the downstream extremely volatile.

**2.2 Model-based measurement of economic uncertainties**

Just like what has been pointed out by Jurado et al. (2015), the existing attempts to measure uncertainties by using proxy variables will lead to significant biases since the fluctuations embedded in these variables may not be tightly linked to economic uncertainties. In their paper, they define the economic uncertainty as the unpredictable components of economy and can be represented by the forecast errors. For example, Rossi & Sekhposyan (2015) echoed this definition and proposed economic uncertainty indices based on distributions of forecast errors drawn from Survey of Professional Forecasters' forecasts. Furthermore, the problem of measuring economic uncertainties has been converted from simply calculating certain indexes into building complex forecasting models. Recent attempts to measuring economic uncertainties by using forecast errors include using factor models (Berger et al., 2016; Mumtaz & Musso, 2021; Mumtaz & Theodoridis, 2017), stochastic volatility models (Berge, 2021; Clark et al., 2020; Jo & Sekkel, 2019), and vector autoregressive (VAR) models (Carriero et al., 2018).

However, the aforementioned attempts may still yield to biased estimation of forecast errors even by using time-varying regression models. That is because true economic conditions are unobservable and key economic indicators may present quite different patterns and inter-correlations. As a result, empirical models with the hidden state assumption will lead to more accurate estimation of forecast errors that can be used to measure economic uncertainties. For example, Bianchi (2016), Bianchi & Melosi (2016; 2017) provide both theoretical and empirical evidence that a Markov-switching model combined with a DSGE setting can be used to better capture the forecast errors of macroeconomic variables. Further applications can also be found in Casarin et al. (2018) and Song & Tang (2022). The HMM model is also widely used in the existing literature to model the hidden states of conditional dynamic processes (Spears, 2021; Zhu et al., 2016). For example, HMM is a quite effective model to detect unobserved financial market status in finance. It has been used to measure the credit risks (Ntwiga et al., 2018) and to predict the debit card transaction frauds (Srivastava et al., 2008). Adopting HMM has also been demonstrated to be an effective way to model key macro-level indicators such as inflation rates (Airaudo & Hajdini, 2021) and business cycles (Barsoum & Stankiewicz, 2015; Guérin

& Marcellino, 2013). Meanwhile, there is also an increasing trend that the researchers choose to use the HMM to analyze the energy market. Take the electricity market as an example, the electricity price dynamics can be decoded and forecasted by the hidden Markov chain (Apergis et al., 2019; Samitas & Armenatzoglou, 2014; Xiong & Mamon, 2019).

In this paper, we follow the idea that true economic regimes are not observable and a Markov-switching mechanism is needed to measure the uncertainties. Firstly, we will use the HMM to assess whether the individual firm's production deviates from its regular pattern after unexpected shocks occur. Secondly, our paper differs from the existing literature that use deviations of forecast errors to measure the economic uncertainties. We propose a new way that the unobserved regimes' probabilities related to the historical observations are used to construct the ECU indices. By comparing regional and sectoral level ECU indexes of Shanghai with other key economic indicators during COVID-19, the ECU indexes that are proposed in this paper have shown to have a great potential to infer the true economic condition uncertainties.

## 3 Research Design

**3.1 Hidden Markov Model**

In a real world, the economic conditions may be prosperous or recessionary and might depend on a variety of factors. Most importantly, true economic condition cannot be readily observed since the impacts of economic condition changes may take a while to manifest. Consequently, changes in economic conditions may not be timely reflected by firms' electricity consumptions. After an unexpected exogenous shock, the firms may continue production for its existing orders. During this time, the electricity consumption remains normal. However, the economic uncertainty enhanced by the lash has led to the latent crisis. This inconsistency in time may also be quite various for different sectors or different regions. Beyond that, a temporary decrease or increase in a firm's electricity consumptions may just be the result of changes in weather conditions or provisional changes of the production schedule. The practical electricity time series cannot be used to describe the economic condition directly, whether it is smoothed or not.

To avoid misjudging the economic conditions, we attempt to recognize the producing behaviors deviation of firms to indicate the economic condition uncertainty. Thus, the empirical strategy should take two factors into consideration to model firms' electricity consumption behavior. Firstly, each firm should select a period of electricity consumption as reference. This reference marks the normal production behavior. By comparing the current electricity consumption to this reference, it can be seen whether production has deviated from firms' usual patterns or not. Secondly, the empirical models should allow the existence of unobserved regimes so that firms could have different production pattern in different economic conditions and the transition between patterns could be explicitly quantified. To meet the above two targets, we choose Hidden Markov Model (HMM) (Baum et al., 1970; Baum & Petrie, 1966) to set up the empirical models which can identify firms' production patterns as well as unobserved economic conditions.

In the HMM setting, we denote the unobservable regimes in period $t$ as $S_t$ and $S_t \in \{S^p, S^r\}$, where $S^p$ stands for prosperous regime and $S^r$ stands for recessionary regime. The regimes is not independent in period $t$; it depends on the regimes in last period and follows a first order Markov chain as follows:

$$\boldsymbol{Q} = \begin{pmatrix} P(S_t = S^p | S_{t-1} = S^p) & P(S_t = S^r | S_{t-1} = S^p) \\ P(S_t = S^p | S_{t-1} = S^r) & P(S_t = S^r | S_{t-1} = S^r) \end{pmatrix} = \begin{pmatrix} q_{pp,t} & q_{pr,t} \\ q_{rp,t} & q_{rr,t} \end{pmatrix} \quad (1)$$

where $\boldsymbol{Q}$ is the transition matrix. Besides, $q_{ij} = P(S_t = S^j | S_{t-1} = S^i)$ ( $i,j \in \{p,r\}$ ) is the conditional probability that $S_t$ equals $S^j$ when $S_{t-1}$ equals $S^i$. The conditional probabilities are called transition probabilities and are assumed to be constant which satisfy the constraints $q_{pp,t} + q_{pr,t} = 1$ and $q_{rp,t} + q_{rr,t} = 1$.

Here we assume that firms' productions show annual periodicity so that we can set the electricity consumption sequence of the last year as the reference. To assess the impact of great natural or economic shocks, the electricity demand in reference year can be regarded as consumption under regular production. By subtracting the firms' electricity consumption of the test sample from that of the reference year, we can obtain a deviation sequence. Let $Y = (y_1, y_2, \cdots, y_T)$ be the observation sequence and $y_t$ be the deviation of 7-day smoothed electricity consumption data at time $t$ ($1 \leq t \leq T$) from the same time in reference year. Then the model specifications for two regimes are as follows:

$$\begin{cases} y_t = \alpha_p t + \beta_p + \varepsilon_{p,t} & if\ S_t = S^p \\ y_t = \alpha_r t + \beta_r + \varepsilon_{r,t} & if\ S_t = S^r \end{cases} \quad (2)$$

where $\alpha$ and $\beta$ are coefficients, and subscripts $p$ and $r$ represent the prosperous and recessionary regime indicator respectively. $\varepsilon_t$ is the i.i.d. normal distribution with mean zero and variance $\sigma^2$, which is conditional on the unobserved regime taking the form like $f_E(\varepsilon_t|S_t = S^i) = \frac{1}{\sqrt{2\pi}\sigma_i} exp\left\{\frac{-\varepsilon_t^2}{2\sigma_i^2}\right\}$. Then the conditional density of $y_t$ is:

$$f(y_t|S_t = S^i) = \frac{1}{\sqrt{2\pi}\sigma_i} exp\left\{\frac{-(y_t - \alpha_i t - \beta_i)^2}{2\sigma_i^2}\right\} \quad (3)$$

Further, we denote the probability matrix corresponding to the above formula as **B**, which is the observation probability matrix:

$$\mathbf{B} = \begin{pmatrix} P(y_t|S_t = S^p) \\ P(y_t|S_t = S^r) \end{pmatrix} = \begin{pmatrix} b_p \\ b_r \end{pmatrix} \quad (4)$$

where $b_i = P(y_t|S_t = S^i)$ is the conditional probability of $y_t$ when $S_t$ equals $S^i$. The unobserved regime $\{S_t\}$ is presumed to be generated by some probability distribution and the unconditional probability that $S_t$ equals to different hidden state is denoted by $\pi_t$. According to the total probability theorem, $\pi_t$ can be estimated from the last $\pi_{t-1}$ and parameter $q_{ij,t-1}$:

$$\mathbf{\Pi}_t = \begin{pmatrix} P(S_t = S^p) \\ P(S_t = S^r) \end{pmatrix}^T = \begin{pmatrix} \pi_{p,t} \\ \pi_{r,t} \end{pmatrix}^T = \begin{pmatrix} q_{pp}\pi_{p,t-1} + q_{rp}\pi_{r,t-1} \\ q_{pr}\pi_{p,t-1} + q_{rr}\pi_{r,t-1} \end{pmatrix}^T = \mathbf{\Pi}_{t-1}\mathbf{P} \quad (5)$$

where $\mathbf{\Pi}_t$ is the observation probability matrix at time $t$. The unconditional probability $\pi_{p,t}$ and $\pi_{r,t}$ are as follows:

$$P(S_t = S^p) = \pi_{p,t}, P(S_t = S^r) = \pi_{r,t} \text{ and } \pi_{p,t} + \pi_{r,t} = 1 \quad (6)$$

Thus, the HMM can be modeled by the **Q**, **B**, $\pi_{p,1}$ and $\pi_{r,1}$. Transition matrix **Q** and initial state probabilities $\pi_{p,1}$ and $\pi_{r,1}$ determine the hidden state sequence. Then the observation probability matrix **B** is used to generate the observation sequence from the hidden state sequence.

Combining formula (1) to (6), then the unconditional density of $y_t$ is the weighted average of all possible values for different regimes:

$$f(y_t) = \frac{\pi_{p,t}}{\sqrt{2\pi}\sigma_p} exp\left\{\frac{-(y_t - \alpha_p t - \beta_p)^2}{2\sigma_p^2}\right\} + \frac{\pi_{r,t}}{\sqrt{2\pi}\sigma_r} exp\left\{\frac{-(y_t - \alpha_r t - \beta_r)^2}{2\sigma_r^2}\right\} \quad (7)$$

All parameters including $\alpha$, $\beta$, $\sigma$, $\pi$ can then be estimated by maximizing the following log likelihood function for all observed data by using EM Algorithm (Bilmes, 1998):

$$L(\alpha, \beta, \sigma, \pi) = \sum_{t=1}^{T} log f(y_t) \quad (8)$$

Thereby we can obtain the estimators $\hat{\alpha}$, $\hat{\beta}$, $\hat{\sigma}$, $\hat{\pi}$ and then work out the value of matrixes **Q** and **B**. In particular, we can infer the probability of the unobserved state $S_t$ given the observations on $y_t$ by a recursive filtering procedure proposed by Hamilton (1989). This recursive filtering procedure takes the prevenient joint conditional probability $P(S_{t-1} = S^i|y_{t-1}, y_{t-2}, \cdots y_0)$ as input and has the output of

$$\hat{\mu}_{i,t} = P(S_t = S^i | y_t, y_{t-1}, \cdots y_0) \tag{9}$$

where $\hat{\mu}_{i,t}$ refers to the probability of unobserved regime $S^i$ at time $t$. Obviously, $\hat{\mu}_{i,t}$ is related to the historical observations. When a firm has already tended to the prosperous or the recessionary regime, its $\hat{\mu}_{r,t}$ will increase if $y_t$ continues to expand this deviation and vice versa. That is, the prosperous or the recessionary regime does not happen overnight but is gradually caused by accumulated deviations. Another advantage of this design is that it can avoid interference caused by sudden actions such as order boom or production equipment maintenance.

In the next step, we follow the applications in Bianchi (2016) and Casarin et al. (2018) to define the two regimes with explicit attributes, i.e., the prosperous regime and the recessionary regime. An illustration from the estimation results of an individual firm can be found in Appendix A.1. Many previous studies have already shown that uncertainties will bring negative effects to real economic activities (Bloom, 2014; Carriero et al., 2018; Mumtaz & Musso, 2021). Therefore, when a firm is experiencing negative shocks then its $y_t$ series should have a dramatic drop at the beginning and be followed by a recovery process. This means $\widehat{\beta_r}$ should be significantly less than zero and $\widehat{\alpha_r}$ should have a positive value. When a firm operates normally then its $y_t$ series should fluctuate around zero and be irrelevant to time trend $t$. This means $\widehat{\beta_p}$ should be close to zero and $\widehat{\alpha_p}$ should not be significantly different from zero.

## 3.2 Economic Condition Uncertainty Index

The results of HMM provide us with the estimated probabilities of two unobserved regimes for each firm in very period. These estimated probabilities are then used to construct the ECU index. The more likely the firms are in the recessionary regime, the larger the economic condition uncertainty will be. Furthermore, daily electricity consumption can represent the production scale and indicate its impact on the market to some extent. Therefore, the ECU index is a weighted average probability of recessionary regime which takes firm's daily electricity consumption data as the weight. The formula to calculate the aggregate ECU index is as follows:

$$ECU_t = \left(\sum_{k=1}^{N} ele_{k,t} \times \hat{\mu}_{k,r,t}\right) / \sum_{k=1}^{N} ele_{k,t} \tag{10}$$

where $\hat{\mu}_{k,r,t}$ is the estimated probability of recessionary regime of firm $k$ in period $t$, and $ele_{k,t}$ is the corresponding electricity consumption. The ECU index is a dimensionless index that ranges from zero to one.

On the one hand, when firms' electricity consumptions deviate from their previous patterns, the probabilities of recessionary regime will increase. When more firms are likely to be in recession, the ECU index will increase. On the other hand, when firms operate in a similar pattern to their past, the probabilities of recessionary regime will be kept at a low level. This will reduce the ECU index. This design means that a firm facing difficulties or bankruptcy does not represent the decline of the industry or market. Only when there is a large-scale high recession risk will the overall uncertainty of the economic condition increase.

Furthermore, if these firms are separated into different groups according to their trades or locations, then we can also define the ECU indexes at sectoral and regional level as follows:

$$ECU_{sector,t} = \left(\sum_{k=1}^{N_{sector}} ele_{k,t} \times \hat{\mu}_{k,r,t}\right) / \sum_{k=1}^{N_{sector}} ele_{k,t} \quad (11)$$

$$ECU_{region,t} = \left(\sum_{k=1}^{N_{region}} ele_{k,t} \times \hat{\mu}_{k,r,t}\right) / \sum_{k=1}^{N_{region}} ele_{k,t} \quad (12)$$

where subscript $sector$ stands for different sectors and subscript $region$ stands for the regions. The regional and sectoral ECU indexes can be interpreted similarly as the aggregate ECU index. They also range from zero to one. Higher index values mean higher uncertainties in regional or sectoral economic conditions. Due to its dimensionless nature, ECU indexes can be used to make consistent comparisons among different regions or sectors.

## 4 Data

This paper uses daily electricity consumption data of 18228 large firms in Shanghai from November 2018 to May 2020[2]. These firms are categorized into different sectors such as manufacturing, real estate, wholesale and retail, finance, hotel, and catering, etc. according to the *Industrial Classification for National Economic Activities* (NBSPRC, 2017). Table 1 summarizes the ratios of different sectors in the sample and the average daily electricity consumption data for each sector.

**Table 1: Summary of the sample data**

| Sector | Ratio | Average daily electricity consumption (kWh) |
|---|---|---|

---

[2] Daily electricity data are provided by State Grid Shanghai Electric Power Company and preprocessed by using 14-day mean average value to interpolate the missing data and outliers. Data is considered to be an outlier when its value exceeds twice the standard deviation beyond the mean value of a 15-day window period.

| | | |
|---|---|---|
| **Primary sectors** | | |
| Agriculture, forestry, husbandry and fishery | 0.34% | 1871.40 |
| **Secondary sectors** | | |
| Mining | 0.09% | 3377.60 |
| Manufacturing | 42.54% | 8654.80 |
| Electricity, heat, gas and water production and supply | 1.13% | 11592.03 |
| Construction | 2.38% | 2661.97 |
| **Tertiary sectors** | | |
| Wholesale and retail | 8.70% | 4454.09 |
| Transportation, storage and postal services | 4.11% | 11335.49 |
| Hotel and catering | 2.33% | 4699.18 |
| Information transmission, software and information technology services | 2.18% | 8734.28 |
| Finance | 1.59% | 7614.70 |
| Real estate | 16.73% | 6331.18 |
| Leasing and business services | 4.03% | 7337.30 |
| Scientific research and technology | 1.49% | 8596.99 |
| Water conservancy, environment and public facilities management | 3.39% | 1940.06 |
| Residential services, repair and other services | 0.35% | 2217.51 |
| Education | 0.78% | 8440.72 |
| Health and social work | 1.54% | 8963.35 |
| Culture, sports and entertainment | 1.49% | 2936.55 |
| Public administration, social security and social organizations | 2.36% | 3274.85 |
| Others | 2.45% | 3669.08 |

Shanghai 's economy is dominated by the secondary and tertiary sectors. In 2019, the primary, secondary and tertiary sectors have accounted for 0.27%, 26.99% and 72.73% of Shanghai's GDP (Shanghai BS, n.d.). In our sample, total number of firms in these three sectors accounted for 0.34%, 46.14% and 53.52% respectively. Among them, manufacturing alone accounted for 42.54%, followed by real estate, which accounted for 16.73%. On the regional level, the economic structures of Shanghai's sixteen districts are quite different. As shown in Fig.1, the downtown areas such as Hongkou, Changning, Putuo, Xuhui, Jingan, Yangpu and Huangpu have very high proportion of tertiary industry. In contrast, the proportion of secondary industry accounts for nearly half or even more in the suburbs, such as Baoshan, Qingpu, Fengxian, Minhang, Jinshan, Pudong, Songjiang,

Jiading, and Chongming.

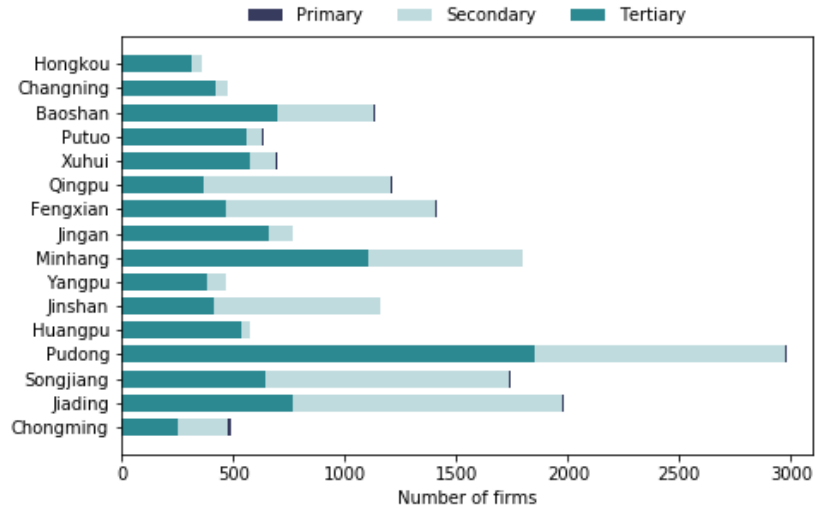

**Fig.1: Distribution of primary, secondary and tertiary sectors**

Since the outbreak of COVID-19 in January 2020, there has been numerous large-scale lockdowns in China. This pandemic has brought a great negative impact on the economy and society, thus we can take the outbreak of COVID-19 as a specific exogenous shock to check whether the ECU indexes can capture the resultant uncertainties in Shanghai's economy. Chinese government extended the length of Spring Festival holiday by three days in order to control the spread of the pandemic[3]. As for Shanghai's local policies, firms were forbidden to resume their businesses until February 9. Some firms even encouraged their employees to work from home for a longer period. Both lockdown and recovery periods have been included in our data sample. Therefore, we are able to estimate the economic disruption of COVID-19 and the follow-up recovery process.

Another crucial factor is that the dates of Spring Festival holiday are not fixed per Gregorian calendar and vary every year. As one of the most important traditional festivals in China, Spring Festival has significant impacts on the country's economy. Before the arrival of this public holiday, firms suspend production activities and almost everything will be put on hold until next year. If this has not been factored in, it is likely to be identified as an anomaly when it is compared with reference year. By considering this issue, we chose the New Year's Eve in 2019 and 2020 as the base points (denoted Day zero) and 95 consecutive days before and after the base points to conduct the HMM estimation. The final periods of the experimental samples included are: from November 1, 2018 to May 10, 2019 (New Year's Eve: February 4, 2019), and from October 21, 2019 to April 28, 2020 (New Year's Eve: January 24, 2020). The electricity consumption time series of the former is taken as the reference, while the latter is a sample of economic condition to be observed. The sequence obtained by subtracting the smoothed reference from the smoothed test samples of each firm is the $y_t$ in HMM.

---

[3] The Spring Festival holiday in 2020 was originally scheduled from 24th to 30th January. Affected by the epidemic, the holiday was later extended to 2nd February.

## 5 Results

### 5.1 Economic Condition Uncertainty index of Shanghai

Fig.2 shows the aggregate ECU index of Shanghai from October 21, 2019 to April 28, 2020. In the fourth quarter of 2019, the overall economic condition of Shanghai was relatively stable, with ECU index maintained at 50.25%. The market did not fluctuate sharply. Since January 14, 2020, the uncertainty of Shanghai's economic condition began to decline gradually. During the Spring Festival holiday (gray shadowed area), aggregate ECU index reached the lowest level, which was about 24% lower than previous average level.

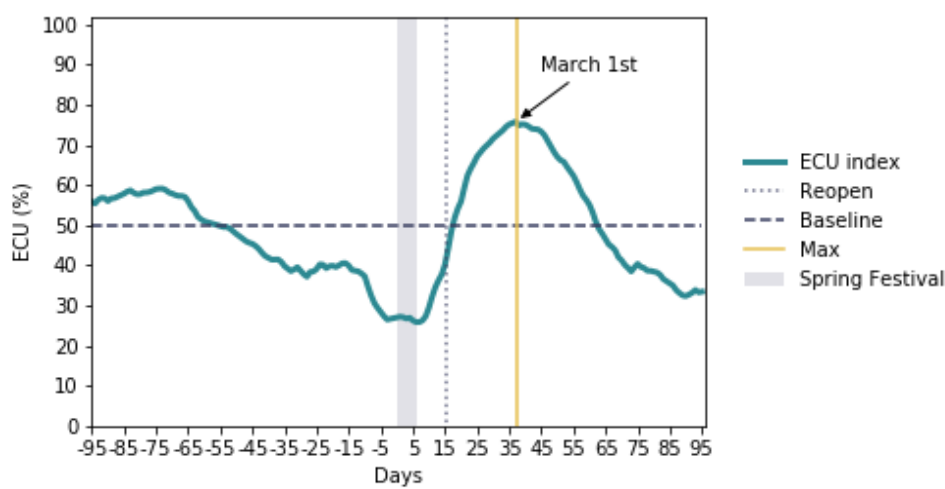

**Fig.2: Aggregate ECU index of Shanghai**

Under normal circumstances, firms will restart production after the Spring Festival holiday. However, due to the implementation of COVID-19 prevention policies, firms' electricity consumption did not follow the previous pattern, that is, the consumption increased after the holiday. This led to the rapid rise of ECU in Shanghai after the Spring Festival holiday. Although firms were allowed to resume production from February 9 (gray dotted line), the whole society was still in fear of COVID-19. Most people were reluctant to go back to office immediately and preferred to work from home, and firms were also not prepared to reopen their businesses. These uncertainties of the economic conditions were directly reflected in the aggregate ECU index, which continued to rise for nearly 22 days after the resumption. The peak ECU index occurred on March 1 (yellow line) and was 25.48% higher than the average level in the fourth quarter in 2019.

In order to restore the overall economic condition as soon as possible, the government had released a series of economic incentive policies like financial subsidies and support, tax reduction, consumer coupons, etc. It took around 25 days for the ECU of Shanghai to drop from the peak to the average level before the outbreak of COVID-19. After the end of March, Shanghai's economy gradually recovered and became more stable. By now the aggregate ECU index was about 13.2% lower than the average level in the fourth quarter in 2019.

**5.2 Sectoral ECU indexes**

Fig.3 illustrates the ECU indexes of primary, secondary and tertiary sectors. The green solid line in the figure represents the aggregate sectoral ECU index and the light green solid lines are sub-sectoral ECU indexes. We can see from Fig.3 that ECU index of primary sector fluctuated greatly before the Spring Festival. The index had two small peaks caused by New Year's Day and weather conditions that reached 70.45% and 59.77%, respectively. In contrast, the aggregate ECU indexes of secondary and tertiary sector were more stable. Similar to the aggregate ECU index of Shanghai, two aggregate sectoral ECU indexes decreased by 10%-20% before the Spring Festival (gray shadowed area), which represented the small uncertainties in economic conditions due to the holiday.

Although all three sectors' aggregate ECU indexes had similar trend that increased first and then decreased to their normal level, they still exhibited different patterns in the duration of COVID-19. Unlike secondary sector that uncertainties decreased quickly after reaching the peak level, tertiary sector had experienced a much longer period of being affected by the COVID-19. There are three main reasons. Firstly, most jobs in tertiary sector, such as finance and IT, are easier to be done from home. As a result, electricity consumptions of firms in this sector were still kept in a relatively low level during the recovery period. Secondly, cinemas, shopping malls, music halls, stadiums and other public venues usually attract large number of people. Therefore, these places were subject to stricter prevention and control regulations for a longer time. Thirdly, the long-lasting travelling bans had greatly affected the tourism and related businesses and the tertiary sector was not surprisingly expected to take a beating. To sum up the above reasons, ECU indexes of tertiary sector were kept at a high level for about 23 days. Also, the average level in this period was 19% higher than that in fourth quarter of 2019. After the Qingming holiday (April 5, 2020), ECU indexes of all three sectors gradually became stable and economic conditions gradually returned to the level before the outbreak of the COVID-19.

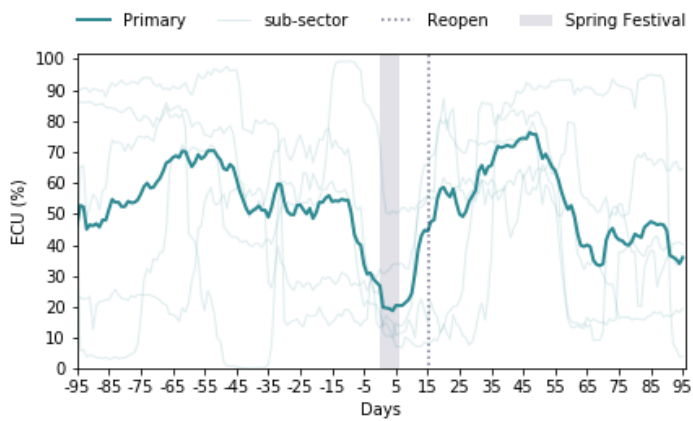

(a) Primary sectors

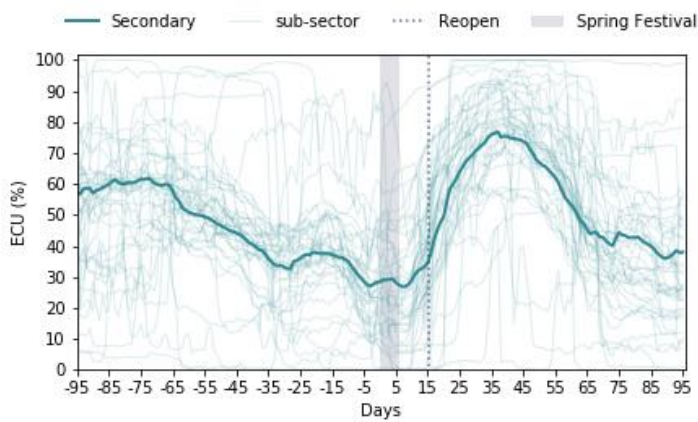

(b) Secondary sectors

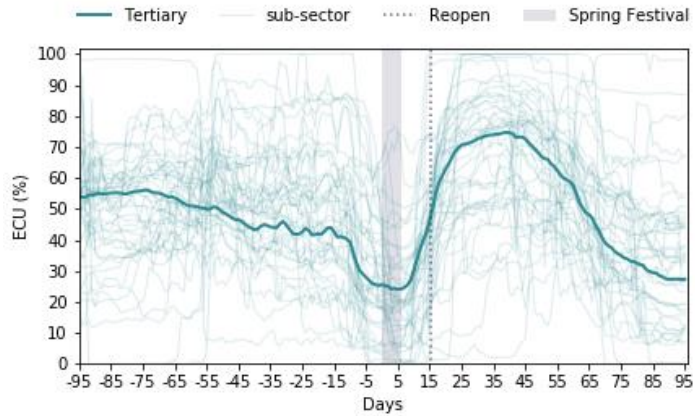

(c) Tertiary sectors

Fig.3: Sector-level ECUs in Shanghai

**5.3 Regional ECU indexes**

As for the Shanghai regional economic condition uncertainty, the overall trends of ECU indexes in the sixteen districts are shown in Fig.4. There were mainly two features of the regional ECU indexes. Firstly, during the pandemic period, tertiary sectors' ECU indexes of several districts were higher than secondary sectors' ECU indexes of these districts. Take Huangpu District as an example. It is in the city proper of Shanghai and is famous for its prosperous business and tourism. Huangpu's tertiary sector's ECU index increased significantly in the early 2020, 35.97% higher than that of the secondary sectors. Secondly, duration of tertiary sectors' ECU indexes with high uncertainty was longer than that of secondary sectors. These results are similar to the results of sectoral aggregate ECU indexes. Shanghai government had imposed stricter restrictions on offline stores, hotels, restaurants and other entertainment venues during the COVID-19. Most tertiary sector services were forced to shut down or put a limit on the number of visitors. As a result, ECU indexes of tertiary sectors in most districts increased earlier than that of secondary sectors. This phenomenon in Jingan District is the most obvious. The peak value of its tertiary sector's ECU index appeared on February 17, 15 days ahead of the peak time of its secondary sector and remained high for nearly a month.

From the above analysis, it can be seen that there are differences in sector conditions among various districts. Considering that the restriction policies were equally implemented in all districts in Shanghai, the heterogeneity of districts' ECU indexes are mainly caused by their economic structure. We find that for districts with higher industrial diversities and more complex economic structures, their ECU indexes would increase less during the COVID-19 period. For example, Minhang District and Pudong District accommodate a variety of industries as well as large numbers of large firms in the secondary sector. Their ECU indexes increased less than that of other districts under the impacts of COVID-19. Moreover, sectoral ECU indexes in these districts were also more stable in general, which means they have more stable economic conditions all the time. For districts with relatively centralized industry or simple economic structures, such as Chongming, the economic conditions were more easily affected by exogenous shocks. Since large-scale farming is not the dominant economic businesses in Shanghai, there is a very limited number of large firms in the primary sector. For the same reason, primary sectors' ECU indexes in all districts presented a pattern of large fluctuations.

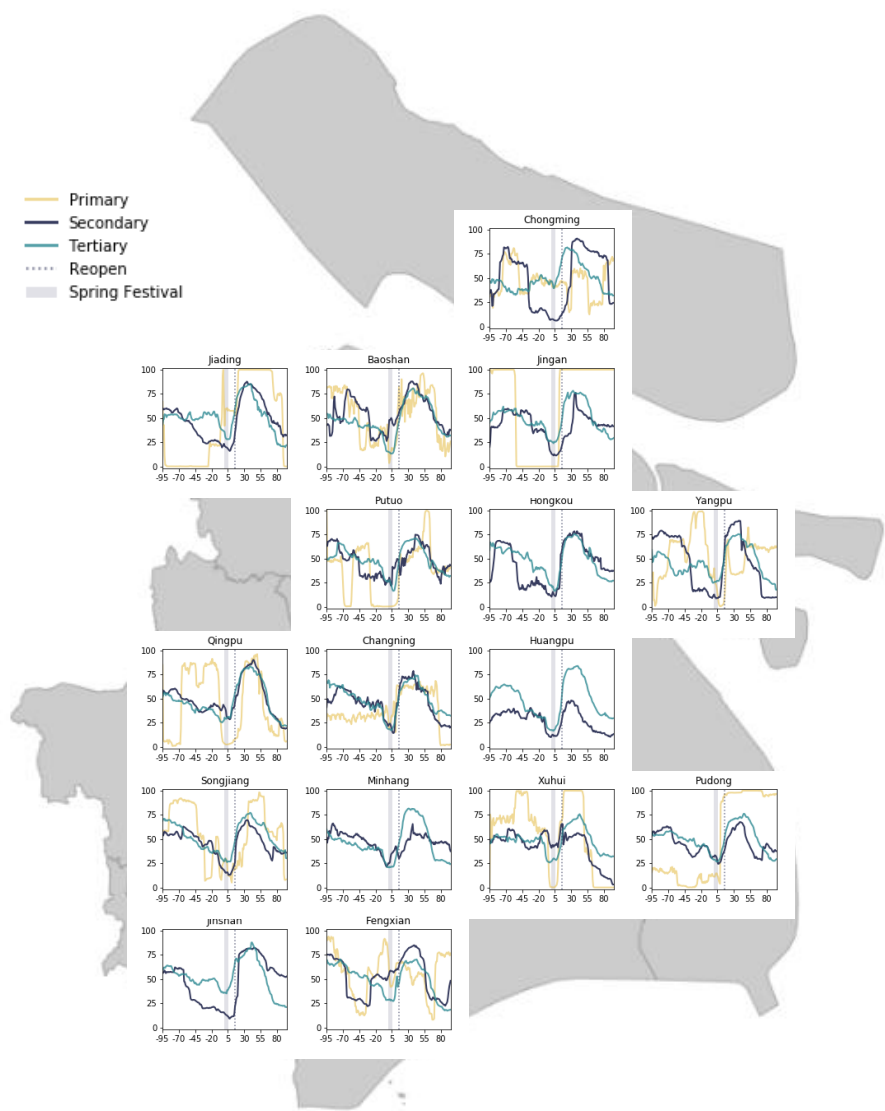

**Fig.4: ECU indexes of 16 districts in Shanghai**

# 6 Discussion

To clarify the meanings of ECU index in characterizing economic conditions, we compare it with four indexes covering electricity consumption, Internet search results, traffic, and finance respectively. These indexes include simplified Resumption Power Index (sRPI), Baidu Search Index, Shanghai Traffic Congestion Index (TCI), Shanghai Interbank Offered Rate (SHIBOR), and can characterize the economic situations from different perspectives to a certain extent. In a similar way to deal with the raw electricity consumption data in ECU, we calculated the 7-day smoothed differences of sRPI, Baidu Search Index and Shanghai TCI in order to eliminate the impacts of Spring Festival and weekends. Since SHIBOR isn't recorded at weekend, we only calculated its differences from that in the reference year.

The sRPI is the total electricity consumption of observed firms with constant coefficients and is derived from the Resumption Power Index (XinhuaNet, 2020), which was widely used to describe the economic resumption after the COVID-19. The Resumption Power Index indicates the difference between the current electricity consumption and that of previous years, rather than the potential uncertainty in the market. As shown in Fig.5(a), △sRPI was negative after the Spring Festival and the sRPI deviated from the reference year. Even though △sRPI rebounded from February 9 (gray dotted line), meaning that Shanghai's economy was recovering from the shock of the epidemic, ECU index still increased for 21 days (from February 9 to March 1, 2020). This indicates that even though the government allows resumption, the market was not confident with the economic conditions in the early stages of economic recovery. Besides, the resumption of each sector was not simultaneous during these days and this is another reason that the uncertainty of economic condition expanded. The manufacturing industry contributed 57.47% to total △sRPI growth, followed by the real estate, while the remaining 17 subsectors account for no more than 30% (Fig.5(b)). In manufacturing, metals and chemicals had the largest increase in △sRPI; and on the contrary, the △sRPI of Petroleum, coke and nuclear fuel processing, etc. even continued to decline (see more details in Appendix A.3). In terms of districts (Fig.5(c)), we can see that the large firms in the suburbs of Shanghai have recovered faster than those in the urban areas. The inconsistency of sectoral and regional recovery will both lead to an increase in the uncertainty of the overall economic conditions, which cannot be revealed by the Resumption Power Index alone.

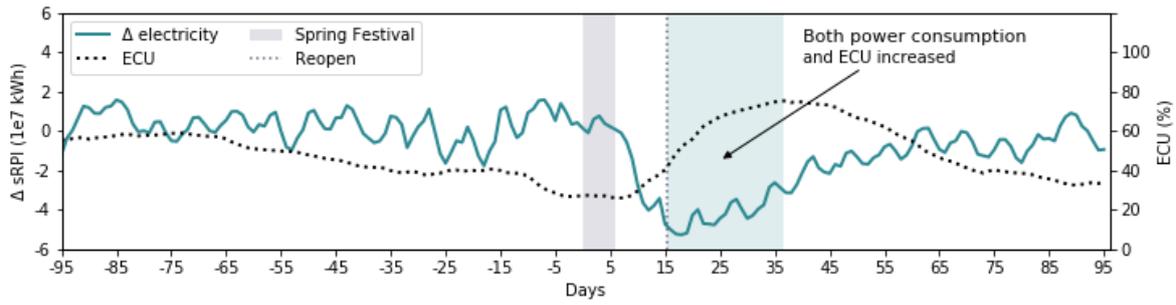

(a) Comparison of ECU and sRPI

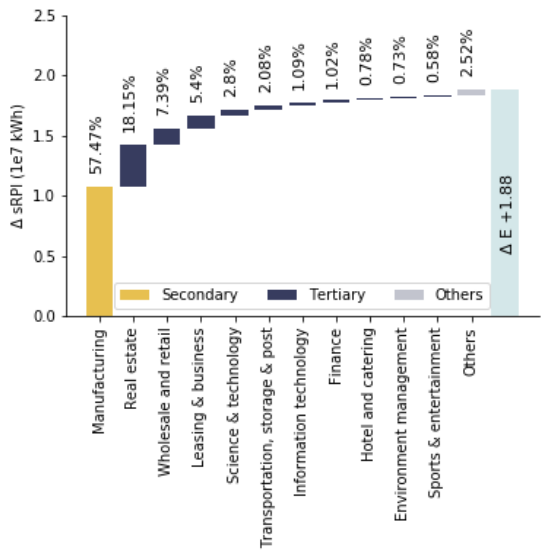      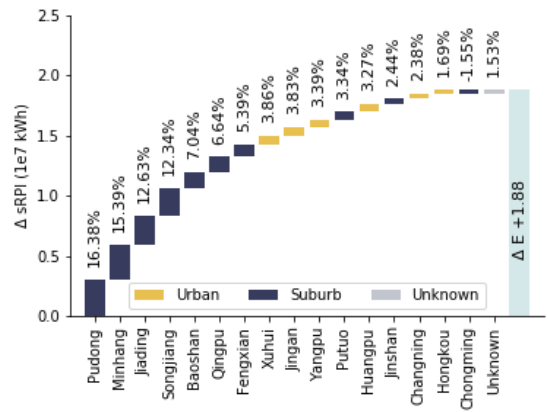

(b) 21-day sRPI growth in terms of sectors     (c) 21-day sRPI growth in terms of districts

Fig.5: ECU and sRPI in Shanghai

Baidu Search Index is a kind of index based on the weighted sum of users' search frequency for a certain word in Baidu search pages. We selected fourteen key words and collected their PC and mobile Baidu Search Index in Shanghai. Generally speaking, the Baidu Search Indexes of the words related to the production grew first. The focus on *Return to Work* and *Work from Home* have already risen sharply before February 9 and their Baidu Search Indexes trended to be stable in mid-March (Fig.6(a)). Due to the increase of the economic condition uncertainty brought by the COVID-19, *Bankrupt* has attracted more hits since March. As for the people's livelihood, *Rent* drew great concern after Spring Festival and its Baidu Search Index continued to be high in the next four months; while *Housing Price*, *Salary* and *Unemployment* all peaked in April (Fig.6(b)). With the uncertainty of the overall economic conditions staying at a high level from the end of February to the beginning of March, the government has subsequently introduced policies to stabilize the economy. Among them, the search heat of *Subsidy* and *Coupon* remained high until May (Fig.6(c)). In addition, the whole society also put the *Economy*, *Foreign Trade* and *Stock Market* on the front burner under the instable economic condition (Fig.6(d)).

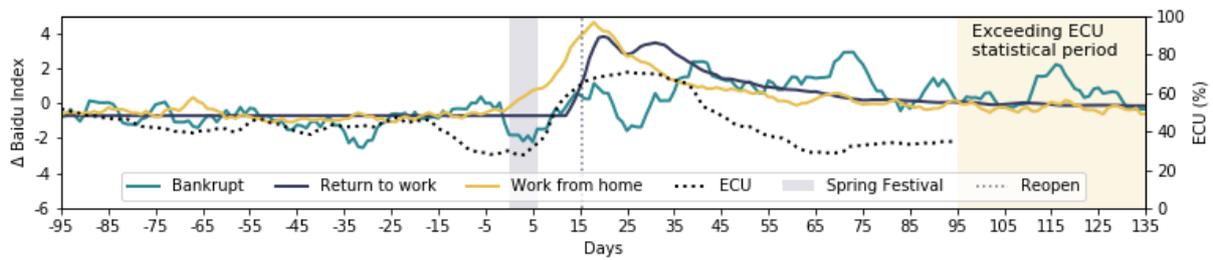

**(a) Baidu Search Indexes on production**

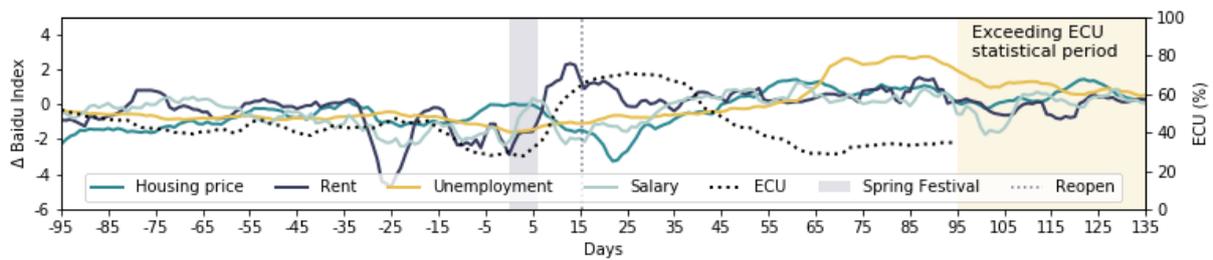

**(b) Baidu Search Indexes on people's livelihood**

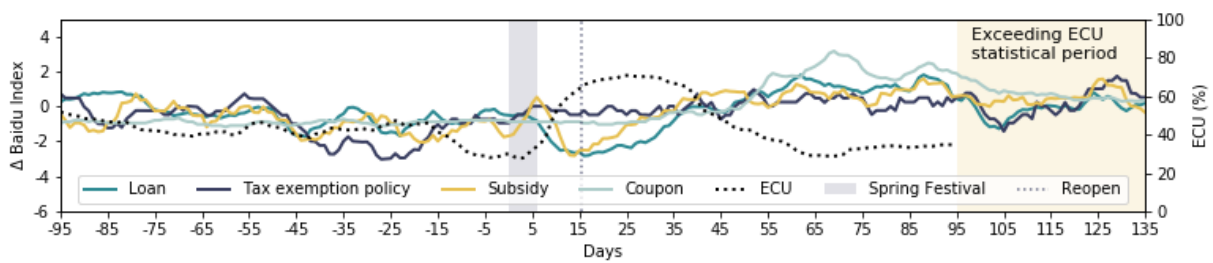

**(c) Baidu Search Indexes on policy**

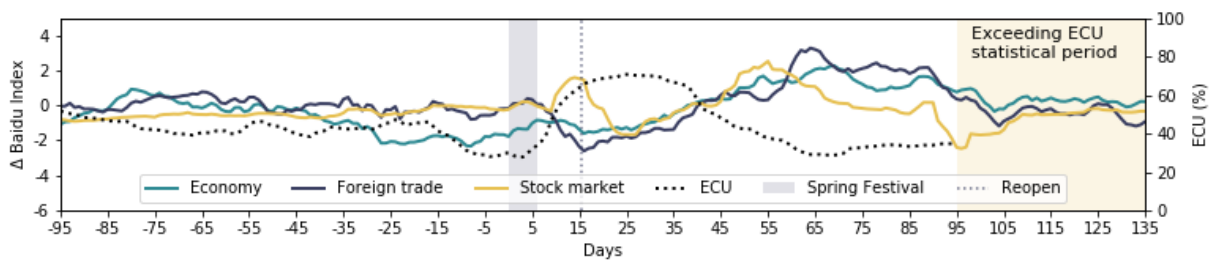

**(d) Other Baidu Search Indexes**

**Fig.6: Baidu Search Indexes in Shanghai**

Shanghai Traffic Congestion Index (TCI) is a set of normalized indexes integrating road design capacity and real-time vehicle speed[4]. It includes 47 elevated observation points and 69 ground road observation points. People's travelling activities and road congestion can reflect the economic vitality indirectly. The less people travel on the roads, the worse the economy is. Thus, we intercept the ΔTCI of elevated and ground road in the morning and evening peak hours (Fig.7). Since Spring Festival, the ΔTCIs in the evening rush hours have already begun to decline. After resuming to work, Shanghai road congestion during 7:00-10:00 and 17:00-20:00 decreased significantly compared to the reference year and the average ΔTCI reduced by about 10. In general, the trends of Shanghai traffic congestion are roughly consistent with ECU index, and both ΔTCIs' valley and ECU's peak appeared about a month after returning to work.

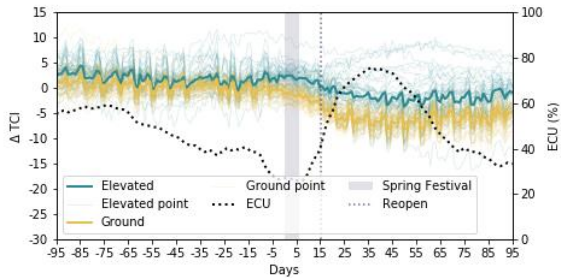

(a) ΔTCI at 7:00 – 8:00

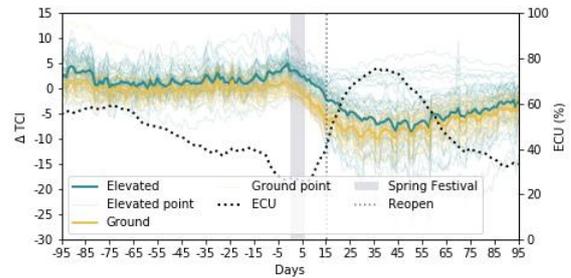

(b) ΔTCI at 17:00 – 18:00

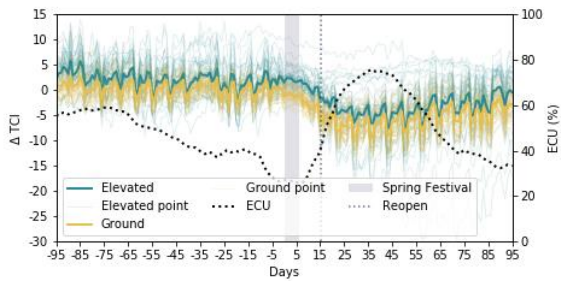

(c) ΔTCI at 8:00 – 9:00

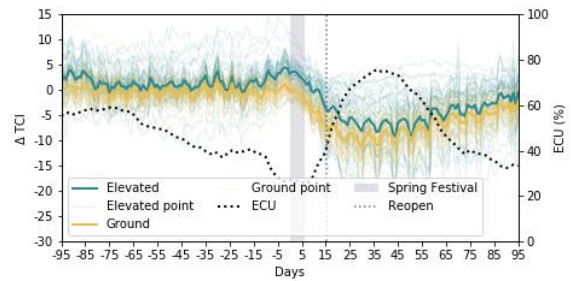

(d) ΔTCI at 18:00 – 19:00

---

[4] Shanghai Traffic Congestion Index is launched by Shanghai Urban and Rural Construction and Transportation Development Research Institute.

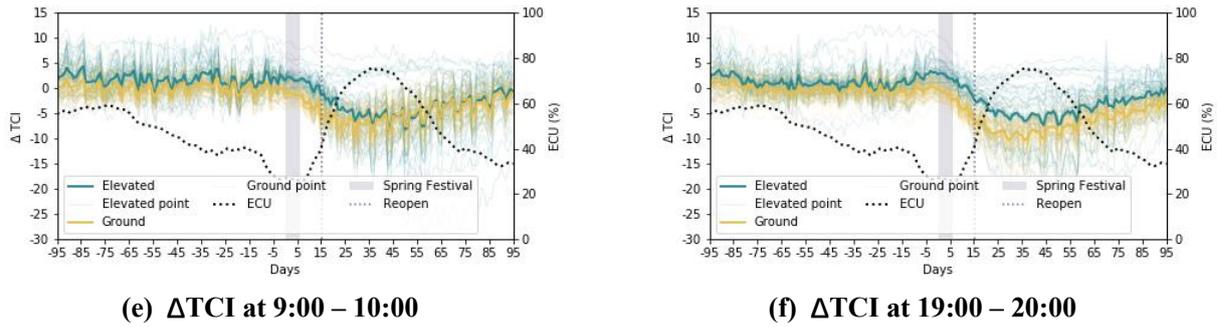

(e) ΔTCI at 9:00 – 10:00    (f) ΔTCI at 19:00 – 20:00

**Fig.7: Traffic Congestion Index in Shanghai**

SHIBOR is the average interest rate determined by the RMB interbank offered rate reported by 18 commercial banks. Fig.8 shows the ΔSHIBORs of overnight, 1 week and 2 weeks (see Appendix A.4 for other SHIBOR indicators). Under the high economic condition uncertainty, all these kinds of varieties are deviated from the reference year and continued to decline while fluctuating after 2020 Spring Festival holiday.

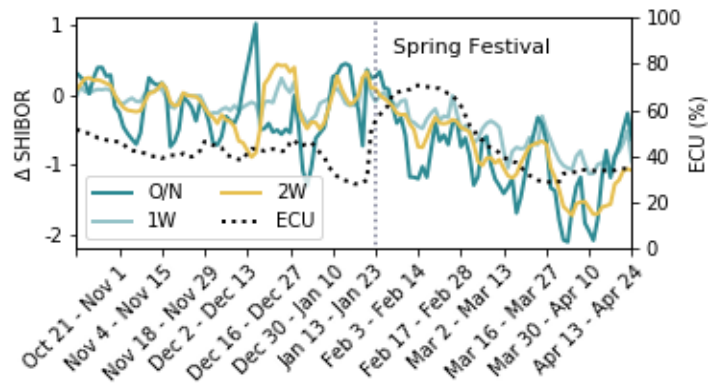

**Fig.8: Shanghai Interbank Offered Rate**

# 7 Conclusions and Policy Implications

## 7.1 Brief summary of the ECU index

There are many examples in both academic and practice fields of people attempting to infer the economic conditions timely (Amisano & Geweke, 2017). However, in the real world, economic conditions can be hardly observed, coupled with the time lag between actual production activities and economic risks, using absolute value of different proxy variables to infer the economic conditions may lead to biases. Moreover, it is the uncertainty of economic conditions, not the absolute level of certain economic condition indexes, plays the core role in decision making process. In this paper, we propose a novel method to construct the ECU index that can be used to infer the economic condition uncertainties. This method uses the hidden Markov process to solve the problem that the switching of macro-economic conditions can be hardly observed. Through focusing on the deviations of electricity consumptions, this method can figure out the prosperous and recessionary states that are used to describe the uncertainties accurately. Then, the ECU index is constructed by the weighted average of all individual firms' probabilities in recessionary economic conditions. The ECU index is a dimensionless index ranges from zero to one, which makes it comparable among sectors, regions and various time periods.

We used the daily electricity consumption data of nearly 20 thousand firms in Shanghai from 2018 to 2020 to construct the ECU indexes of Shanghai. Results show that all ECU indexes, whether at sectoral level or regional level, have successfully captured the negative impacts of COVID-19 on Shanghai's economic conditions. Besides, the ECU indexes also presented the heterogeneities in different districts as well as in different sectors. This demonstrates that changes in uncertainties of economic conditions are mainly related to regional economic structures and targeted regulatory policies faced by sectors. Districts with more complex economic structures typically suffered less from COVID-19 and had relatively stable and low uncertainties of economic conditions. From the sector's perspective, tertiary sector faced strictest regulatory policies and businesses in this sector were the easiest to move to work-from-home mode. Consequently, ECU index of the tertiary sector typically rose higher and lasted longer (see Appendix A.2). In addition, we explained the highly uncertain period of Shanghai's economic conditions from the aspects of electricity consumption, network search, traffic, and finance.

**7.2 Policy implications**

The ECU index has been demonstrated to be capable in inferring economic condition uncertainties. So, it can be seen as an eligible candidate to be used in decision making process to understand the risks and conditions of the economy. Moreover, the ECU index proposed in this paper can be extended in two ways to facilitate the decision making. Firstly, the method based on HMM estimation can be extended to include more than two hidden statuses. Then a set of indexes can be constructed to reflect those more complicated economic conditions. This extension will be useful when decision-making procedures are sensitive to true economic conditions. Secondly, the ECU index can be constructed by other data apart from electricity consumption. All data that have high frequencies and present regular periodic patterns can be used to construct different ECU indexes. For example, custom data can be used to construct the ECU index to measure uncertainties in international trade; macro financial data can be used to construct the ECU index to measure uncertainties in financial market; etc. These extensions are not subject to a specific region or field, which provides a huge potential for future research.

# Acknowledgement

This work was supported by the Key Program of National Philosophy and Social Science Foundation Grant [grant no 15ZDB148], the National Natural Science Foundation of China [grant nos 72234002, 71925010, 72234001], and the Shanghai Talent Development Fund [2021098]. We thank State Grid Shanghai Electric Power Company for providing access to the data used in this study; and Fudan Association for Computational Social Science for providing technical supports.

# Appendix A

**Appendix A.1 Illustration of a sample firm's estimation results**

Fig.A1 shows the electricity consumption behavior characteristics of one firm in 191 days. Fig.A1(a) describes the differences of electricity consumptions between two periods before and after smoothing. It can be seen that differences fluctuated around zero before the Spring Festival holiday. Differences were significantly negative due to the lockdown policies within two months after the Spring Festival holiday. Fig.A.1(b) shows the probabilities of recessionary condition (dark blue line) and prosperous condition (yellow line) estimated by HMM. Obviously, the model can accurately identify the abnormal production behavior during the lockdown period in 2020. During this period, the probabilities of recessionary condition were close to one. The probabilities of prosperous condition were close to zero as well. These results are consistent to the trends shown in Fig.A1(a).

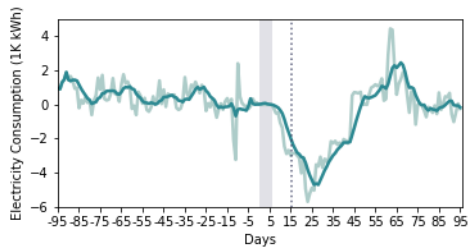 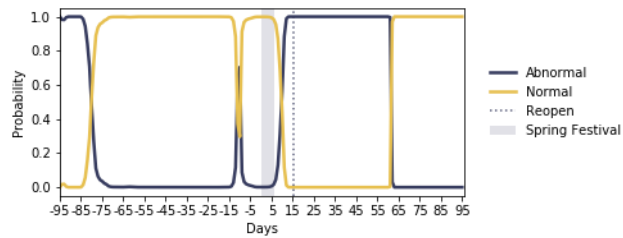

(a) Differences of electricity consumptions  (b) Probabilities of two unobserved economic conditions

**Fig.A1: Electricity consumption data and estimated probabilities of a sample firm**

**Appendix A.2 Major subsectors' ECU indexes in Shanghai**

Fig.A2 shows ECU indexes (green lines) of nine major subsectors in Shanghai, including manufacturing, construction, wholesale and retail, hotel & catering, finance, real estate, scientific research & technology, transportation & storage & postal services, and education. The Spring Festival holiday (gray shadowed area) and reopen time (gray dotted line) are marked in each subfigure. Sample sizes and maximum ECU indexes for each subsector are shown in subfigures.

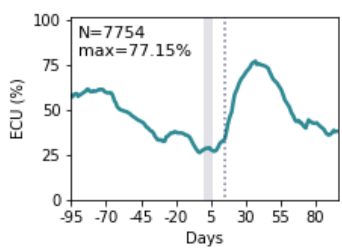 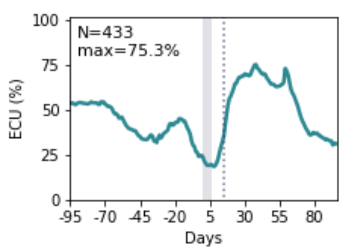 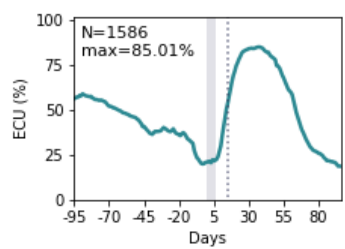

(a) Manufacturing  (b) Construction  (c) Wholesale and retail

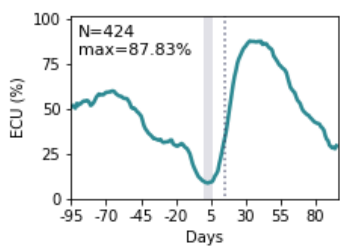 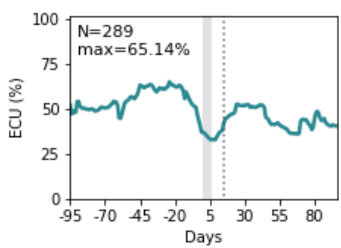 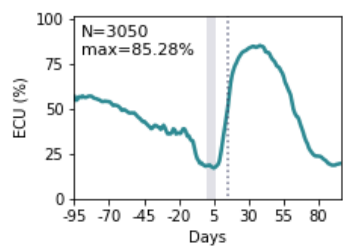

(d) Hotel and catering  (e) Finance  (f) Real estate

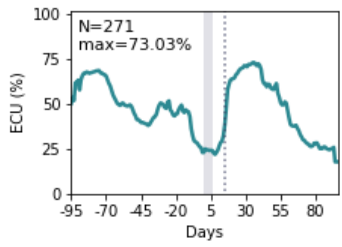 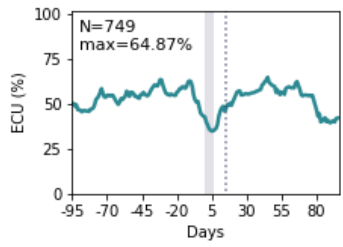 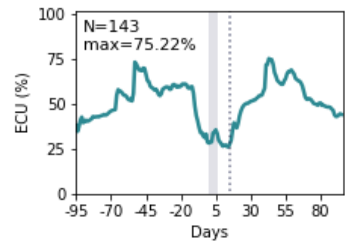

(g) Scientific research and technology

(h) Transportation, storage and postal services

(i) Education

Fig.A2: ECU indexes of major subsectors

## Appendix A.3 Simplified RPI growth in manufacturing

Manufacturing is the biggest contributor to the increase of electricity consumption during the ECU growth period. Fig.A3 shows the composition of 21-day sRPI growth in manufacturing. The figure analyzes the $\Delta$sRPI growth of 31 items (dark blue bar) and their proportion in the overall growth of manufacturing in these 21 days.

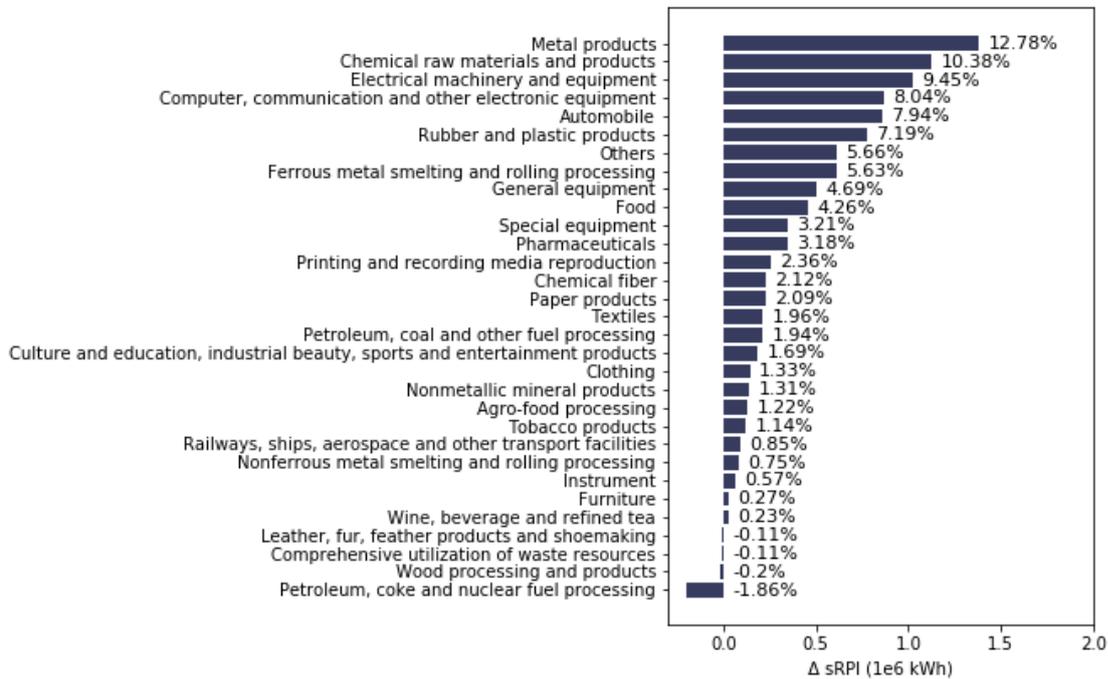

Fig.A3: Composition of 21-day sRPI growth in manufacturing

## Appendix A.4 Other SHIBOR varieties

Fig.A4 shows the 5-day (green line), 10-day (dark blue line) and 20-day (yellow line) means of overnight ΔSHIBORs. It can be seen that the longer the time span, the smoother the SHIBOR curve. All of above declined after Spring Festival.

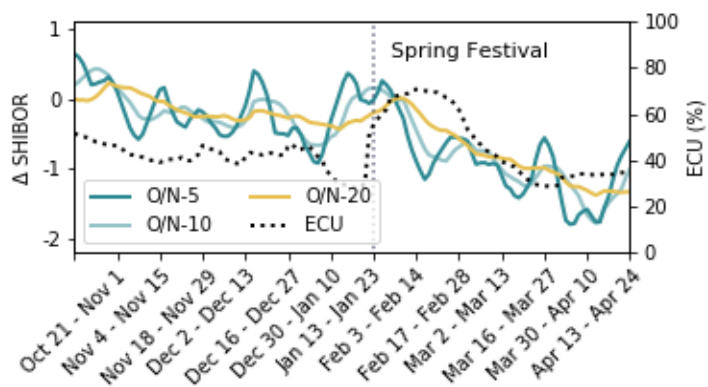

**Fig.A4: Other SHIBOR varieties**

# References

Ahir, H., Bloom, N., & Furceri, D. (2018). The World Uncertainty Index. *SSRN Electronic Journal*. https://doi.org/10.2139/ssrn.3275033

Ahir, H., Bloom, N., & Furceri, D. (2019, September 9). New Index Tracks Trade Uncertainty Across the Globe. *IMF Blog*. https://blogs.imf.org/2019/09/09/new-index-tracks-trade-uncertainty-across-the-globe/

Airaudo, M., & Hajdini, I. (2021). Consistent Expectations Equilibria in Markov Regime Switching Models and Inflation Dynamics. *International Economic Review*, *62*(4), 1401–1430. https://doi.org/10.1111/iere.12529

Amisano, G., & Geweke, J. (2017). Prediction Using Several Macroeconomic Models. *The Review of Economics and Statistics*, *99*(5), 912–925. https://doi.org/10.1162/REST_a_00655

Apergis, N., Gozgor, G., Lau, C. K. M., & Wang, S. (2019). Decoding the Australian electricity market: New evidence from three-regime hidden semi-Markov model. *Energy Economics*, *78*, 129–142. https://doi.org/10.1016/j.eneco.2018.10.038

Arias, M. A., Gascon, C. S., & Rapach, D. E. (2016). Metro business cycles. *Journal of Urban Economics*, *94*, 90–108. https://doi.org/10.1016/j.jue.2016.05.005

Azqueta-Gavaldón, A. (2017). Developing news-based economic policy uncertainty index with unsupervised machine learning. *Economics Letters*, *158*, 47–50.

Bah, M. M., & Azam, M. (2017). Investigating the relationship between electricity consumption and economic growth: Evidence from South Africa. *Renewable and Sustainable Energy Reviews*,

*80*, 531–537. https://doi.org/10.1016/j.rser.2017.05.251

Baker, S. R., Bloom, N., & Davis, S. J. (2016). Measuring economic policy uncertainty. *The Quarterly Journal of Economics*, *131*(4), 1593–1636.

Barsoum, F., & Stankiewicz, S. (2015). Forecasting GDP growth using mixed-frequency models with switching regimes. *International Journal of Forecasting*, *31*(1), 33–50. https://doi.org/10.1016/j.ijforecast.2014.04.002

Baum, L. E., & Petrie, T. (1966). Statistical Inference for Probabilistic Functions of Finite State Markov Chains. *The Annals of Mathematical Statistics*, *37*(6), 1554–1563.

Baum, L. E., Petrie, T., Soules, G., & Weiss, N. (1970). A Maximization Technique Occurring in the Statistical Analysis of Probabilistic Functions of Markov Chains. *The Annals of Mathematical Statistics*, *41*(1), 164–171.

Berge, T. J. (2021). Time-Varying Uncertainty of the Federal Reserve's Output Gap Estimate. *The Review of Economics and Statistics*, 1–38. https://doi.org/10.1162/rest_a_01102

Berger, T., Grabert, S., & Kempa, B. (2016). Global and Country-Specific Output Growth Uncertainty and Macroeconomic Performance. *Oxford Bulletin of Economics and Statistics*, *78*(5), 694–716. https://doi.org/10.1111/obes.12118

Beyer, R. C. M., Franco-Bedoya, S., & Galdo, V. (2021). Examining the economic impact of COVID-19 in India through daily electricity consumption and nighttime light intensity. *World Development*, *140*, 105287. https://doi.org/10.1016/j.worlddev.2020.105287

Bianchi, F. (2016). Methods for measuring expectations and uncertainty in Markov-switching models. *Journal of Econometrics*, *190*(1), 79–99. https://doi.org/10.1016/j.jeconom.2015.08.004

Bianchi, F., & Melosi, L. (2016). Modeling the Evolution of Expectations and Uncertainty in General Equilibrium. *International Economic Review*, *57*(2), 717–756. https://doi.org/10.1111/iere.12174

Bianchi, F., & Melosi, L. (2017). Escaping the Great Recession. *American Economic Review*, *107*(4), 1030–1058. https://doi.org/10.1257/aer.20160186

Bilmes, J. A. (1998). A Gentle Tutorial of the EM Algorithm and its Application to Parameter Estimation for Gaussian Mixture and Hidden Markov Models. *International Computer Science Institute*, *4*(510), 126.

Bloom, N. (2014). Fluctuations in Uncertainty. *Journal of Economic Perspectives*, *28*(2), 153–176. https://doi.org/10.1257/jep.28.2.153

Bok, B., Caratelli, D., Giannone, D., Sbordone, A. M., & Tambalotti, A. (2018). Macroeconomic Nowcasting and Forecasting with Big Data. *Annual Review of Economics*, *10*(1), 615–643. https://doi.org/10.1146/annurev-economics-080217-053214

Caldara, D., & Iacoviello, M. (2018). Measuring Geopolitical Risk. *Board of Governors of the Federal Reserve Board*, *2019*(1222). https://doi.org/10.17016/IFDP.2018.1222

Carriero, A., Clark, T. E., & Marcellino, M. (2018). Measuring Uncertainty and Its Impact on the Economy. *The Review of Economics and Statistics*, *100*(5), 799–815. https://doi.org/10.1162/rest_a_00693

Carvalho, M., Delgado, D. B. de M., Lima, K. M. de, Cancela, M. de C., Siqueira, C. A. dos, & Souza, D. L. B. de. (2021). Effects of the COVID-19 pandemic on the Brazilian electricity consumption patterns. *International Journal of Energy Research*, *45*(2), 3358–3364. https://doi.org/10.1002/er.5877

Casarin, R., Foroni, C., Marcellino, M., & Ravazzolo, F. (2018). Uncertainty through the lenses of a mixed-frequency Bayesian panel Markov-switching model. *The Annals of Applied Statistics*, *12*(4), 2559–2586. https://doi.org/10.1214/18-AOAS1168

Chetty, R., Friedman, J. N., Hendren, N., & Stepner, M. (2020). *The Economic Impacts of COVID-19: Evidence from a New Public Database Built Using Private Sector Data*. 108.

Clark, T. E., McCracken, M. W., & Mertens, E. (2020). Modeling Time-Varying Uncertainty of Multiple-Horizon Forecast Errors. *The Review of Economics and Statistics*, *102*(1), 17–33. https://doi.org/10.1162/rest_a_00809

de Lange, M., Wolbers, M. H. J., Gesthuizen, M., & Ultee, W. C. (2014). The Impact of Macro- and Micro-Economic Uncertainty on Family Formation in The Netherlands. *European Journal of Population / Revue Européenne de Démographie*, *30*, 161–185. https://doi.org/10.1007/s10680-013-9306-5

Donaldson, D., & Storeygard, A. (2016). The View from Above: Applications of Satellite Data in Economics. *Journal of Economic Perspectives*, *30*(4), 171–198. https://doi.org/10.1257/jep.30.4.171

Dong, L., Chen, S., Cheng, Y., Wu, Z., Li, C., & Wu, H. (2017). Measuring economic activity in China with mobile big data. *EPJ Data Science*, *6*(1), 29. https://doi.org/10.1140/epjds/s13688-017-0125-5

FED. (n.d.). *Economic Conditions Index for St. Louis, MO-IL (MSA) (DISCONTINUED)*. FRED, Federal Reserve Bank of St. Louis; FRED, Federal Reserve Bank of St. Louis. Retrieved June 25, 2021, from https://fred.stlouisfed.org/series/STLAGRIDX

Gibson, J., Olivia, S., & Boe-Gibson, G. (2020). Night Lights in Economics: Sources and Uses. *Journal of Economic Surveys*, *34*(5), 955–980. https://doi.org/10.1111/joes.12387

Gibson, J., Olivia, S., Boe-Gibson, G., & Li, C. (2021). Which night lights data should we use in economics, and where? *Journal of Development Economics*, *149*, 102602. https://doi.org/10.1016/j.jdeveco.2020.102602

Goel, R. K., & Ram, R. (2013). Economic uncertainty and corruption: Evidence from a large cross-country data set. *Applied Economics*, *45*(24), 3462–3468. https://doi.org/10.1080/00036846.2012.714073

Guérin, P., & Marcellino, M. (2013). Markov-Switching MIDAS Models. *Journal of Business & Economic Statistics*, *31*(1), 45–56. https://doi.org/10.1080/07350015.2012.727721

Hamilton, J. D. (1989). A New Approach to the Economic Analysis of Nonstationary Time Series and the Business Cycle. *Econometrica*, *57*(2), 357–384. https://doi.org/10.2307/1912559

Henderson, J. V., Storeygard, A., & Weil, D. N. (2012). Measuring Economic Growth from Outer Space. *American Economic Review*, *102*(2), 994–1028. https://doi.org/10.1257/aer.102.2.994

Holm, J. R., & Østergaard, C. R. (2015). Regional Employment Growth, Shocks and Regional Industrial Resilience: A Quantitative Analysis of the Danish ICT Sector. *Regional Studies*, *49*(1), 95–112. https://doi.org/10.1080/00343404.2013.787159

Huang, Y., & Luk, P. (2020). Measuring economic policy uncertainty in China. *China Economic Review*, *59*, 101367.

IMF. (n.d.). *World Uncertainty Index*. World Uncertainty Index. Retrieved August 14, 2021, from https://worlduncertaintyindex.com/

Jean, N., Burke, M., Xie, M., Davis, W. M., Lobell, D. B., & Ermon, S. (2016). Combining satellite imagery and machine learning to predict poverty. *Science*, *353*(6301), 790–794. https://doi.org/10.1126/science.aaf7894

Jo, S., & Sekkel, R. (2019). Macroeconomic Uncertainty Through the Lens of Professional Forecasters. *Journal of Business & Economic Statistics*, *37*(3), 436–446. https://doi.org/10.1080/07350015.2017.1356729

Jurado, K., Ludvigson, S. C., & Ng, S. (2015). Measuring Uncertainty. *American Economic Review*, *105*(3), 1177–1216. https://doi.org/10.1257/aer.20131193

Koop, G., McIntyre, S., Mitchell, J., & Poon, A. (2020). Regional output growth in the United Kingdom: More timely and higher frequency estimates from 1970. *Journal of Applied Econometrics*, *35*(2), 176–197. https://doi.org/10.1002/jae.2748

Krarti, M., & Aldubyan, M. (2021). Review analysis of COVID-19 impact on electricity demand for residential buildings. *Renewable and Sustainable Energy Reviews*, *143*, 110888. https://doi.org/10.1016/j.rser.2021.110888

Lasarte-López, J. M., Carbonero-Ruz, M., Nekhay, O., & Rodero-Cosano, M. L. (2020). Why do regional economies behave differently? A modelling approach to analyse region-specific dynamics along the business cycle. *Applied Economics Letters*, *0*(0), 1–9. https://doi.org/10.1080/13504851.2020.1820436

Lee, C.-C., Wang, C.-W., Ho, S.-J., & Wu, T.-P. (2021). The impact of natural disaster on energy consumption: International evidence. *Energy Economics*, *97*, 105021. https://doi.org/10.1016/j.eneco.2020.105021

Liu, D., Ruan, L., Liu, J., Huan, H., Zhang, G., Feng, Y., & Li, Y. (2018). Electricity consumption and economic growth nexus in Beijing: A causal analysis of quarterly sectoral data. *Renewable and Sustainable Energy Reviews*, *82*, 2498–2503. https://doi.org/10.1016/j.rser.2017.09.016

Markhvida, M., Walsh, B., Hallegatte, S., & Baker, J. (2020). Quantification of disaster impacts through household well-being losses. *Nature Sustainability*, *3*(7), 538–547. https://doi.org/10.1038/s41893-020-0508-7

Mumtaz, H., & Musso, A. (2021). The Evolving Impact of Global, Region-Specific, and Country-Specific Uncertainty. *Journal of Business & Economic Statistics*, *39*(2), 466–481. https://doi.org/10.1080/07350015.2019.1668798

Mumtaz, H., & Theodoridis, K. (2017). Common and country specific economic uncertainty. *Journal of International Economics*, *105*, 205–216. https://doi.org/10.1016/j.jinteco.2017.01.007

NBSPRC. (n.d.). *China Macro-economic Prosperity Index*. Retrieved August 14, 2021, from https://data.cnki.net/StatisticFocus/Article?id=12

NBSPRC. (2017, June 30). *Industrial classification for national economic activities*. National Bureau of Statistics. http://www.stats.gov.cn/tjsj/tjbz/hyflbz/201710/t20171012_1541679.html

Ng, S., & Wright, J. H. (2013). Facts and Challenges from the Great Recession for Forecasting and Macroeconomic Modeling. *Journal of Economic Literature*, *51*(4), 1120–1154. https://doi.org/10.1257/jel.51.4.1120

Norbutas, L., & Corten, R. (2018). Network structure and economic prosperity in municipalities: A large-scale test of social capital theory using social media data. *Social Networks*, *52*, 120–134. https://doi.org/10.1016/j.socnet.2017.06.002

Ntwiga, D., Ogutu, C., Kiura, M., & Weke, P. (2018). A Hidden Markov Model of Risk Classification among the Low Income Earners. *Journal of Finance and Economics*, *6*, 242–249. https://doi.org/10.12691/jfe-6-6-6

Obschonka, M., Stuetzer, M., Audretsch, D. B., Rentfrow, P. J., Potter, J., & Gosling, S. D. (2016). Macropsychological Factors Predict Regional Economic Resilience During a Major Economic Crisis. *Social Psychological and Personality Science*, *7*(2), 95–104. https://doi.org/10.1177/1948550615608402

PBC. (n.d.). *Prosperity index of 5000 enterprises*. Retrieved August 17, 2021, from https://data.cnki.net/StatisticFocus/Article?id=31

Peters, B. G. (2021). Governing in a time of global crises: The good, the bad, and the merely normal. *Global Public Policy and Governance*, *1*(1), 4–19. https://doi.org/10.1007/s43508-021-00006-x

Rossi, B., & Sekhposyan, T. (2015). Macroeconomic Uncertainty Indices Based on Nowcast and Forecast Error Distributions. *American Economic Review*, *105*(5), 650–655. https://doi.org/10.1257/aer.p20151124

Samitas, A., & Armenatzoglou, A. (2014). Regression tree model versus Markov regime switching: A comparison for electricity spot price modelling and forecasting. *Operational Research*, *14*(3), 319–340. https://doi.org/10.1007/s12351-014-0149-6

Sensier, M., Bristow, G., & Healy, A. (2016). Measuring Regional Economic Resilience across Europe: Operationalizing a complex concept. *Spatial Economic Analysis*, *11*(2), 128–151. https://doi.org/10.1080/17421772.2016.1129435

Shahbaz, M., Sarwar, S., Chen, W., & Malik, M. N. (2017). Dynamics of electricity consumption, oil price and economic growth: Global perspective. *Energy Policy*, *108*, 256–270. https://doi.org/10.1016/j.enpol.2017.06.006

Shahbaz, M., Sbia, R., Hamdi, H., & Ozturk, I. (2014). Economic growth, electricity consumption, urbanization and environmental degradation relationship in United Arab Emirates. *Ecological Indicators*, *45*, 622–631. https://doi.org/10.1016/j.ecolind.2014.05.022

Shanghai BS. (n.d.). *Shanghai Statistical Yearbook 2020*. Shanghai Bureau of Statistics. Retrieved June 26, 2021, from

http://tjj.sh.gov.cn/tjnj/20210303/2abf188275224739bd5bce9bf128aca8.html

Shin, M., Zhang, B., Zhong, M., & Lee, D. J. (2018). Measuring international uncertainty: The case of Korea. *Economics Letters*, *162*, 22–26. https://doi.org/10.1016/j.econlet.2017.10.014

Song, D., & Tang, J. (2022). News-Driven Uncertainty Fluctuations. *Journal of Business & Economic Statistics*, *0*(ja), 1–35. https://doi.org/10.1080/07350015.2022.2097912

Song, J. (2018). *Analysis of Regional Economic Development Differences and Causes in Guangdong Province*. 531–537. https://doi.org/10.2991/hssmee-18.2018.90

Spears, R. (2021). The Impact of Public Opinion on Large Global Companies' Market Valuations: A Markov Switching Model Approach. *Journal of Finance and Economics*, *9*(3), 115–141. https://doi.org/10.12691/jfe-9-3-3

Srivastava, A., Kundu, A., Sural, S., & Majumdar, A. (2008). Credit Card Fraud Detection Using Hidden Markov Model. *IEEE Transactions on Dependable and Secure Computing*, *5*(1), 37–48. https://doi.org/10.1109/TDSC.2007.70228

Sun, S., & Anwar, S. (2015). Electricity consumption, industrial production, and entrepreneurship in Singapore. *Energy Policy*, *77*, 70–78. https://doi.org/10.1016/j.enpol.2014.11.036

Tierney, K. (2012). Disaster Governance: Social, Political, and Economic Dimensions. *Annual Review of Environment and Resources*, *37*(1), 341–363. https://doi.org/10.1146/annurev-environ-020911-095618

WB. (n.d.). *World Development Indicators*. Retrieved August 12, 2021, from https://datatopics.worldbank.org/world-development-indicators/

XinhuaNet. (2020, February 21). *E. China's Zhejiang uses big data to aid work resumption*. http://www.xinhuanet.com/english/2020-02/21/c_138805551.htm

Xiong, H., & Mamon, R. (2019). A higher-order Markov chain-modulated model for electricity spot-price dynamics. *Applied Energy*, *233–234*, 495–515. https://doi.org/10.1016/j.apenergy.2018.09.039

Yalta, A. T. (2011). Analyzing energy consumption and GDP nexus using maximum entropy bootstrap: The case of Turkey. *Energy Economics*, *33*(3), 453–460. https://doi.org/10.1016/j.eneco.2010.12.005

Yu, Y., & Du, Y. (2019). Impact of technological innovation on CO2 emissions and emissions trend prediction on 'New Normal' economy in China. *Atmospheric Pollution Research*, *10*(1), 152–161. https://doi.org/10.1016/j.apr.2018.07.005

Zhang, C., Zhou, K., Yang, S., & Shao, Z. (2017). On electricity consumption and economic growth in China. *Renewable and Sustainable Energy Reviews*, *76*, 353–368. https://doi.org/10.1016/j.rser.2017.03.071

Zhou, L., & Chen, Z. (2021). Are CGE models reliable for disaster impact analyses? *Economic Systems Research*, *33*(1), 20–46. https://doi.org/10.1080/09535314.2020.1780566

Zhou, Y., Zhang, S., Wu, L., & Tian, Y. (2019). Predicting sectoral electricity consumption based on complex network analysis. *Applied Energy*, *255*, 113790. https://doi.org/10.1016/j.apenergy.2019.113790

Zhu, G., Song, K., Zhang, P., & Wang, L. (2016). A traffic flow state transition model for urban road network based on Hidden Markov Model. *Neurocomputing*, *214*, 567–574. https://doi.org/10.1016/j.neucom.2016.06.044